\documentclass[twocolumn,showpacs,amsmath,amssymb,prb,graphics,graphicx
]{revtex4-1}
\usepackage{bm}
\usepackage[pdftex]{graphicx}
\usepackage{dcolumn}
\begin{document}

\title{ Homes scaling and BCS}
\author{V. G. Kogan  }
            \affiliation{
               Ames Laboratory - DOE, Ames, IA 50011}
 \begin{abstract}
It is argued on the basis of the BCS theory that the zero-$T$ penetration depth satisfies $ \lambda^{-2}(0)\propto\sigma T_c $ ($\sigma$ is the normal state dc conductivity)  not only in the extreme dirty limit $\xi_0/\ell \gg 1$, but in a broad range of scattering parameters down  to  $\xi_0/\ell \sim 1$ ($\xi_0$ is the zero-$T$ BCS coherence length and $\ell$ is the mean-free path).  Hence, the scaling $ \lambda^{-2}(0)\propto\sigma T_c $, suggested as a new universal property of superconductors,\cite{Homes} finds a natural explanation within the BCS theory.  
\end{abstract}
 \date{ \today}
\maketitle

It has recently been found that the zero-$T$ penetration depth in many superconductors satisfies a scaling relation  $ \lambda^{-2}(0)\propto\sigma T_c $ ($\sigma$ is the normal state dc conductivity) over many orders of magnitude of $  \sigma T_c $.\cite{Homes0,Homes}  A number of non-trivial theoretical ideas were offered to explain this scaling.\cite{Zaanen,Tallon,Basov,Imry,Erd} Here,  standard isotropic BCS superconductors are shown to satisfy this relation in a broad domain of scattering parameters $\xi_0/\ell  $   from the dirty limit $\xi_0/\ell \gg 1 $ down  to  $\xi_0/\ell \sim 1$.

In  isotropic BCS superconductors the penetration depth is given by:
 \begin{equation}
 \lambda ^{-2}= \frac{16\pi^2 e^2TN(0)\Delta ^2v^2
 }{3c^2}\,  \sum_{\omega>0}  \frac{1
}{\beta ^{2}\beta^\prime}  \,. 
 \label{lambda-tensor1}
\end{equation}
Here, $\hbar\omega=\pi T(2n+1)$ defines the Matsubara frequencies, $N(0)$ is the density of states for one spin, $v$ is the Fermi velocity, $\beta^2 = \hbar^2\omega^2+\Delta^2 $, $\beta^\prime=\beta+\hbar/2\tau$, and $\tau$ is the transport scattering time. One can find this result on the last page of the book
by Abrikosov, Gor'kov and Dzyaloshinskii. \cite{3authors}  It can be readily derived   using Eilenberger quasi-classical version of the BCS theory, see, e.g., Refs.\,\onlinecite{nonloc,PK-ROPP}.
 
At zero temperature, one can replace the sum with an integral according to $2\pi T\sum_n \to \int_0^\infty d(\hbar\omega)$ to obtain after straightforward algebra:
 \begin{eqnarray}
 \lambda ^{-2}(0)= \frac{4\pi^2  e^2N(0) 
v^2}{3c^2\eta} \left(1 +\frac{4\tan^{-1}\frac{ \eta-1 }{\sqrt{1-\eta^2} }}{\pi\sqrt{1-\eta^2}}\right)   
 ,\qquad
\label{lambda0}
\end{eqnarray}
where the scattering parameter
\begin{equation}
\eta=\frac{\hbar}{2\tau\Delta_0} = \frac{\pi}{2}\,\frac{\xi_0}{\ell} ,  
\label{eta}
\end{equation}
  Eq.\,(\ref{lambda0}) works for any $\eta>0$. For $\eta>1$, it could be written in explicitly real form by replacing $\tan^{-1}\to -\tanh^{-1}$ and $\sqrt{1-\eta^2} \to \sqrt{ \eta^2-1}$.

In the dirty limit, the scattering parameter $\eta\gg 1$ and one obtains
 \begin{eqnarray}
 \lambda_d ^{-2}(0)= \frac{4\pi^2  e^2 N(0) 
v^2}{3c^2\eta}  = \frac{4\pi^2\sigma\Delta_0}{\hbar c^2} \,  
\label{lambda0-dirty}
\end{eqnarray}
where $\sigma=2e^2N(0)v^2\tau/3$ is the normal state conductivity. This  also follows from the known dirty limit expression:
\begin{equation}
 \lambda_d ^{-2}= \frac{8\pi^2 e^2 N(0)v^2 \tau}{3
c^2\hbar}\,  \Delta  \tanh\frac{\Delta 
}{ 2T}  \,.  \label{lambda-dirty}
\end{equation}

Since $\Delta_0\propto T_c$, Eq.\,(\ref{lambda0-dirty}) prompted suggestions   that the scaling $  \lambda^{-2}(0)\propto\sigma T_c $   can be explained by  strong scattering present in many materials.\cite{Homes-dirty,Tallon} This argument, however, was criticized since the scaling in question seems to work not only for dirty materials.\cite{Homes}

 The question remains, however, how strong the scattering should be for the dirty limit scaling to work. To answer this question one observes that the pre-factor in Eq.\,(\ref{lambda0}) coincides with Eq.\,(\ref{lambda0-dirty}) of the dirty limit, albeit with an arbitrary scattering parameter $\eta$. We denote this pre-factor as $ \lambda_d ^{-2}(0,\eta)$ to avoid confusion with the dirty limit $\eta\to \infty$. Eq.\,(\ref{lambda0}) takes the form:
 \begin{eqnarray}
 \lambda ^{-2}(0)=  \lambda_d ^{-2}(0,\eta)\left(1 +\frac{4\tan^{-1}\frac{ \eta-1 }{\sqrt{1-\eta^2} }}{\pi\sqrt{1-\eta^2}}\right).  
\label{lambda0a}
\end{eqnarray}

\begin{figure}[h]
\begin{center}
 \includegraphics[width=8cm] {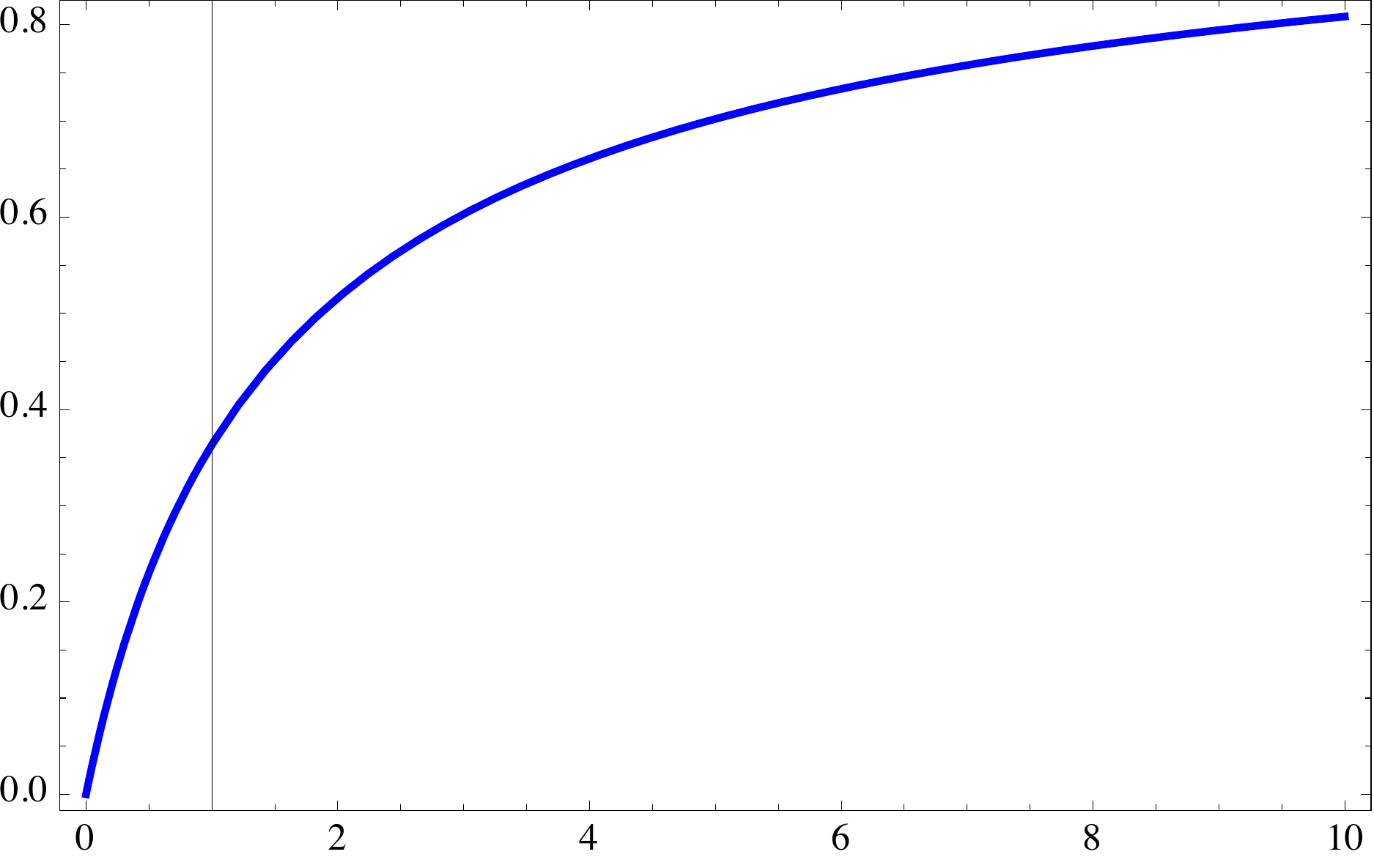}
\caption{Parentheses of Eq.\,(\ref{lambda0a}) versus $\eta $ for $0<\eta<10$.  
}
\label{fig1}
\end{center}
\end{figure}

Note that $ \lambda_d ^{-2}(0,\eta)=4\pi^2\sigma\Delta_0/\hbar c^2\propto\sigma T_c$, the same scaling as in the dirty limit. Hence, deviations from this scaling are determined by the expression in parentheses. 
Fig.\,\ref{fig1} shows that this expression varies only by a factor of 2 when the scattering parameter $\eta$ changes from 10 to 1, the latter value corresponding to the quite clean situation with $\xi_0/\ell =2/\pi=0.64$. 
This suggests that the dirty limit scaling may work quite well in a broad domain of scattering parameters;  even more so visually if one employs log-log plots. 
 
\begin{figure}[h]
\begin{center}
 \includegraphics[width=6cm] {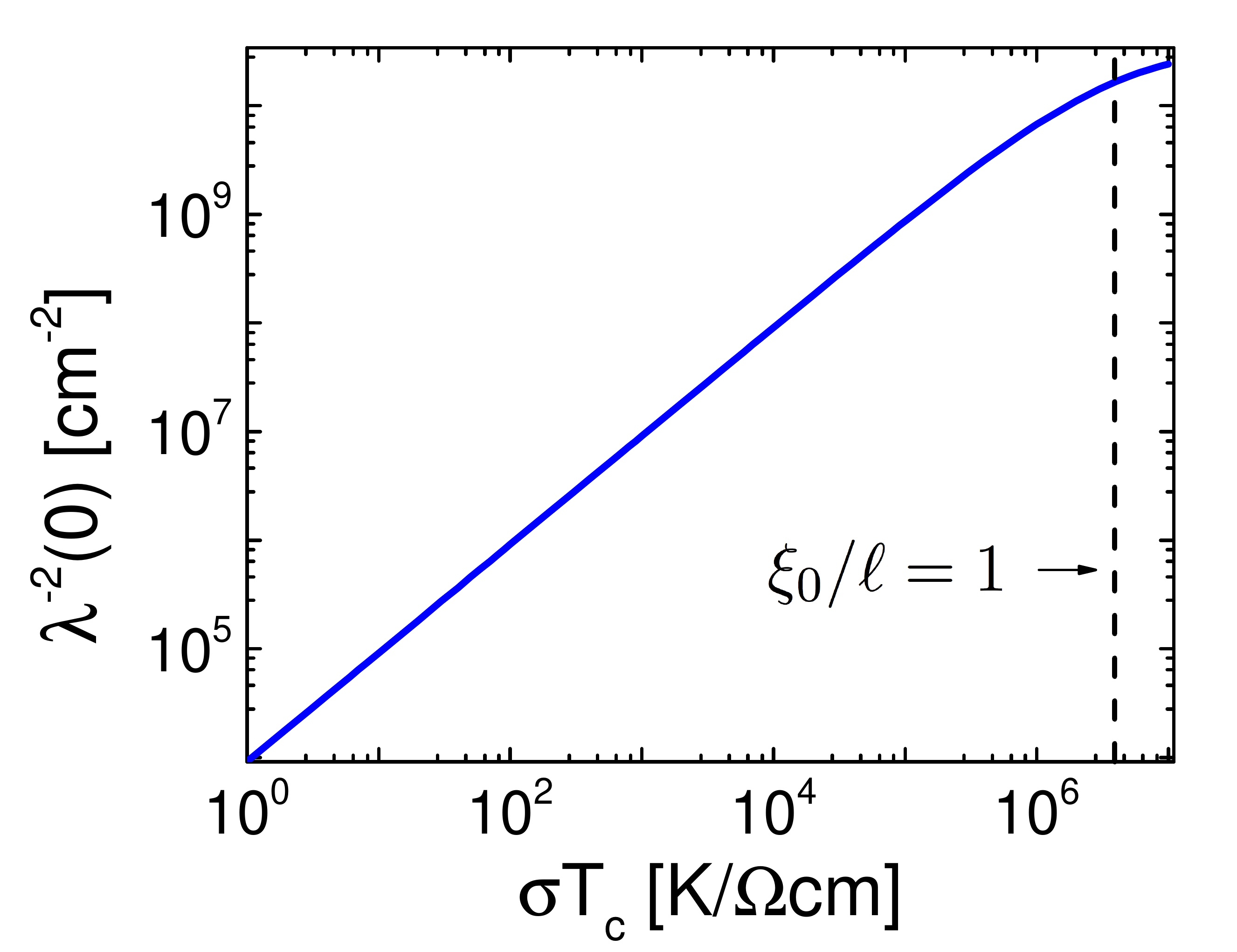}
\caption{The log-log plot of $\lambda^{-2}\, $cm$^{-2}$ versus $x= \sigma T_c$ in K/$\Omega\,$cm.  The vertical line shows $x$ corresponding to $\xi_0/\ell= 1$.}
\label{fig2}
\end{center}
\end{figure}

To show this, we  express $\lambda_d ^{-2}(0,\eta)$ and $\eta$ in terms of the product $x=\sigma T_c$ in K/$\Omega\,$cm since these units are preferred by experimentalists: \cite{Homes}   
 \begin{eqnarray}
 \lambda_d ^{-2}(0,\eta)\,{\rm cm}^{-2}=   0.915\times 10^4\,x\, \nonumber\\
\eta=\frac{4\pi^2 e^2N(0)v^2}{3c^2 \lambda_d ^{-2}(0)}  = \frac{6.1\times 10^6}{ x}\,,   
\label{lam2-dirty}
\end{eqnarray}
Here,  $N(0)v^2=3n/2m\approx 1.65\times 10^{49}$\,CGS for the free electrons  is taken as  an estimate ($n\approx 10^{22}\,$cm$^{-3}$ and $m$ is the free electron mass).  With these numbers Eq.\,(\ref{lambda0a}) generates the curve shown in Fig.\,\ref{fig2}.
 
The left part of this plot corresponds to large scattering parameters $\eta$, whereas the right one represents the clean situation. The boundary between these extremes is $\xi_0/\ell\approx 1$ or $\eta\approx\pi/2$. With the numbers chosen, this corresponds to $  \sigma T_c \approx 3.9\times 10^6\,$K/$\Omega\,$cm. Hence, the figure  shows that in a broad range  of the variable $ \sigma T_c $, the behavior of $\lambda^{-2}(0)$ is in fact  close to that of the dirty limit. The maximum $  \sigma T_c = 10^7\,$K/$\Omega\,$cm of the figure (and of the data collection of Ref.\,\onlinecite{Homes}) corresponds to $\xi_0/\ell \approx 0.2$, i.e., to   clean materials. When the material is in  the  clean limit $\eta\to 0$, the curve of Fig.\,\ref{fig2} flattens to approach 
 \begin{eqnarray}
 \lambda ^{-2}_{\rm clean}(0)= \frac{8\pi   e^2N(0) 
v^2}{3c^2 } =  \frac{4\pi e^2n}{mc^2 } \,.
\label{lambda0cl}
\end{eqnarray}
This, however, happens at very large values of  $ \sigma T_c$ out of the range of available data.\cite{Homes} At the maximum available $  \sigma T_c = 10^7\,$K/$\Omega\,$cm the deviation of the curve on the log-log plot of Fig.\,\ref{fig2} from the   straight line is about 7\%.

Thus, qualitatively, ``Homes scaling", shown in Fig.\,2 of Ref.\,\onlinecite{Homes}, is well reproduced by the BCS theory and does not necessarily call for exotic constructions for its justification.\cite{Zaanen,Imry,Erd} The oversimplified scheme presented here, of course, can be improved by taking into account anisotropies, variations in densities of states, Fermi velocities, pair breaking, {\it etc}. It strongly suggests, however, that the idea  of the dirty limit scaling is certainly viable and can be extended to a  broad range of scattering parameters. The extensive set of data summarized by the Homes scaling can be considered as yet another confirmation of the BCS theory, if any is still needed.\\
 
 The author is grateful to S. Bud'ko, P. Canfield, R.~Prozorov, J. Clem,   V.~Taufour, and H. Kim   for  interest and help. Discussions with C. Homes were welcome and encouraging. The Ames Laboratory is supported by the Department of Energy, Office of  Basic Energy Sciences, Division of Materials Sciences and Engineering under Contract No. DE-AC02-07CH11358.

            \end{document}